# Semiclassical Approximation with Two Small Parameters

N.N.Trunov[1]

*D.I.Mendeleyev Institute for Metrology*

*Russia, St.Peterburg. 190005 Moskovsky pr. 19*

(Dated: December 25, 2009)

**Abstract:** We propose a new variant of the semiclassical quantization with two independent parameters. The first one is proportional to the Planck constant as usually and the second one is connected with a deviation of the given potential from a very wide and important class of them for which an exact condition can be constructed. The accuracy is estimated for a number of special cases.

## 1. Introduction

The semiclassical approximation with all its variations is still one of the main, simplest and most universal methods for solving problems where the exact solutions are cumbersome or unknown – not only in the quantum theory but also in various fields of science. Advantages of these methods as compared with the computer's numerical ones are not only their simplicity but also a clear physical meaning of each stage of calculation. In particular, the explicit dependence on a given potential and other parameters can be easily traced. In its turn, it allows to introduce new physical conceptions and parameters. For example, an effective quantum number for centrally symmetric problem was constructed [1] and used for explaining the structure of the Periodic system of elements [1, 2].

---

[1] Electronic address: trunov@vniim.ru

Hereafter we propose a variant with two parameters, the first of them is the Planck constant (or the dimensionless inverse number of bound states). The second parameter indicated the deviation of a given potential from the special, but very wide and important class of potentials, for which an exact semiclassical condition is constructed. Specific forms of such parameter may depend on studied problems. Our approach can simplify or make more precise many calculations.

## 2. The semiclassical series expansion

Without loss of generality we can write the equation determining energy levels $\varepsilon_n$ in a potential well $V(x)$ in the following form:

$$\Phi(\varepsilon) = \frac{1}{\pi\beta} \int \sqrt{\varepsilon - V} \, dx = n + \frac{1}{2} + \delta \tag{1}$$

with the unknown yet function $\delta$ which ensures the exact spectrum. Here $n = 0, 1, 2...$, the integral is taken between turning points and

$$\beta^2 = \hbar^2 / 2m \tag{2}$$

The common popular semiclassical condition corresponds to $\delta \equiv 0$. It is known that such condition is exact only for few potentials. A power series expansion

$$\delta = \sum_{k=1}^{\infty} \delta_k$$
$$\delta_k = f_k(\varepsilon) \beta^{2k-1} \tag{3}$$

was formally built for $\delta$ [3] but is not practically used since $\delta_k$ are very cumbersome for $k > 1$,

$$\delta_1 = \frac{\beta}{24\pi}\frac{\partial^2}{\partial\varepsilon^2}\int\frac{dx}{\sqrt{\varepsilon-V}}\left(\frac{dV}{dx}\right)^2. \tag{4}$$

## 3. A class of potentials for which the exact explicit semiclassical condition exists

For all potentials which may be expressed by means of an auxiliary function $s(x)$ as [4]:

$$V(x) = A^2 s^2 + Bs + C$$
$$\sigma \equiv \frac{ds}{dx} = a_2 s^2 + a_1 s + a_0 \tag{5}$$

the quantization condition (1) is exact if we put [6]

$$\delta = \frac{2\delta_1}{1+\sqrt{1+16\delta_1^2}}; \tag{6}$$

It should be stressed that (5) is not a given ad hoc class but it embraces all the potentials used in handbooks as model and reference ones for the semiclassical approximation as well as for exact solutions [5]. (Excepting those potentials for which spectra are roots of some transcendental functions). This class embraces the potentials related to the factorization method [6] as well as to the supersymmetric theory with an additive parameter [7].

For all the potentials (5) and only for them

$$\gamma(\varepsilon) = \frac{d\delta_1}{d\varepsilon} \equiv 0, \tag{7}$$

so that $\delta$ (6) is also independent on $\varepsilon$. The value of $\delta_1$ (4),

$$\delta_1 = \frac{\beta a_2}{8A} \tag{8}$$

is invariant under the transformation $V \to V - C$, $s \to s + const$, in many cases it is convenient to choose $V = A^2 s^2$ (with the same $a_2, A$).

We can also obtain from (6) all higher corrections as a series expansion in powers of $\delta_1$:

$$\delta_3 = -4\delta_1^3 \qquad (9)$$

and so on for all the potentials (5). Though such expansion is impossible if $16\delta_1^2 \geq 1$, our $\delta$ (6) remains exact at any $\delta_1$. The maximum absolute value $|\delta| = 1/2$ is reached when $|\delta_1| \to \infty$.

In a general case each $\delta_n$ as well as $\delta$ itself may be expressed as some functional of $\delta_1(\varepsilon)$ and $\Phi(\varepsilon)$. It follows from the fact proved earlier [8]: each potential is fully determined by $\delta_1(\varepsilon)$ and $\Phi(\varepsilon)$. Thus the same pair $[\delta_1(\varepsilon), \Phi(\varepsilon)]$ determines the spectrum and in particular $\delta(\varepsilon)$.

Dependence of $\delta_n(\varepsilon)$ on $\delta_1(\varepsilon)$ and $\Phi(\varepsilon)$ may be certainly very cumbersome. But one may expect that simple approximate expressions exist for the potentials which are similar to (5).

## 4. Semiclassical approximations with two parameters

Now let's suppose that our potential $V(r)$ almost belongs to the class (5) so that a small deviation may be characterized by a small parameter $t$. We shall specify $t$ for each subclass of similar potentials, e.g. $t$ may be chosen proportional to $\gamma$ (7).

We can develop a variant of the semiclassical approximation with two small parameters: $\beta$ and $t$. For example, for the standard WKB case we have:

$$\Phi(\varepsilon) = n + \frac{1}{2} + O(\beta) \qquad (10)$$

where $O(\beta)$ demonstrates the order of omitted terms. If we take into account $\delta_1(\varepsilon)$, we obtain

$$\Phi(\varepsilon) = n + \frac{1}{2} + O(\beta^3) \tag{11}$$

If we include $\delta(\varepsilon)$ (6) as a whole,

$$\Phi(\varepsilon) = n + \frac{1}{2} + \delta(\varepsilon) + O(\beta^3 t) \tag{12}$$

Really, developing $\delta(\varepsilon)$ (6) into the series we obtain the linear term $\delta_1$ which is valid for all potentials and the third in $\beta$ order term $-4\delta_1^3(\varepsilon)$. This term is exact if $t = 0$, see (9), so that it may deviate as

$$\delta_3(\varepsilon) = -4\delta_1^3(\varepsilon)\left[1 + tf(\varepsilon)\right] \tag{13}$$

with some function $f(\varepsilon) = O(1)$. As it was discussed above, $f$ is a functional of $\Phi$.

Since the whole set of the model and reference potentials (5) exhausts (or almost exhausts) all the interesting types of potentials the deviation of the latter from (5) may be described by a small parameter $t$.

## 5. The Sturmian problem

In this section we demonstrate our two-parametric approach for observing the appearance of new bound states by increasing the strength of our potential well with a sole parabolic minimum $V = 0$ and asymptotics

$$V(-\infty) = W, \quad V(\infty) = U, \tag{14}$$

$$r^2 = W/U \geq 1 \tag{15}$$

If $V$ (14) belongs to the class (5), its parameters are

$$\begin{aligned} V &= Us^2 \\ \sigma(s) &= (1+s)(r-s) \end{aligned} \tag{16}$$

Really, limiting values when $\sigma = 0$ are $r = s$ and $s = -1$; corresponding values of $V$ coincide with (14).

According to (8), since $a_2 = -1$,

$$\delta_1 = -\frac{\beta}{8\sqrt{U}} \tag{17}$$

independently on $W$. When a new bound state appears, its energy $\varepsilon = U$. For such a case, introducing a new variable $s$ so that $dx = ds/\sigma(s)$ we obtain

$$\Phi(U) = \frac{\sqrt{U}}{\beta} k(r) \tag{18}$$

$$k(r) = 1 - \sqrt{\frac{r-1}{r+1}} \tag{19}$$

with $r$ (15). Note that if we fix $W$ and change $U$, $k(r)$ will depend on $U$; we can also fix $k$, then $W$ increases together with $U$.

Comparing (17) and (18) we obtain the following condition:

$$\delta_1 \Phi(U) = -\frac{1}{8k(r)} \tag{20}$$

for all the potentials (5).

Now if we have a potential with a small deviation from the class (5) – it manifests as a weak dependence $\delta$ on $\varepsilon$, so that in (7) $\gamma \neq 0$ - we can use (20) for a simple approximate expression determining values of $U$ at which new bound states appear:

$$\Phi(U) = n + \frac{1}{2} - \frac{1}{8\Phi(U)k(r)} + O(\beta t) + O(\beta^3). \tag{21}$$

As it is clear from (20) and (21), the genuine dimensionless small parameter proportional to $\beta$ is

$$b = \frac{1}{8(N - 1/2)}, \tag{22}$$

where $N$ is the total number of bound states ($N = 1$ if $n = 0$).

An important advantage of the above condition (21) is as follows. We have avoided an extra job to calculate $\delta_1$. Since $\delta_1$ (4) contains many differentiations it requires the exact analytic form of $V(x)$, unlike $\Phi$ (1), which can be calculated with $V(x)$ given numerically.

If we nevertheless have calculated $\delta(U)$ independently on $\Phi(U)$, we can use $\delta(U)$ in order to improve the exactness of our condition and at the same time to exclude the condition $N \gg 1$ (practically $N \geq 3$), which requires for the smallness of $b$ (22).

Let's find on the basic of $\delta$ (6) an approximate expression for $\delta$, which is valid for all $N$ and has the following properties:

a) only known functions $\delta(U)$, $\Phi(U)$, $k(r)$ can be used;

b) $\delta$ remains valid for all $V$ (5), (when $\delta_1$ does not depend on $\varepsilon$);

c) $\delta = \delta_1 + O(\beta^3)$ at any $V$ if $|\delta_1| \ll 1$;

d) this new condition should be valid in an opposite extreme case $\Phi(U) \to 0$ and $k=1$. It is known that the lowest bound state with $n = 0$ does exist if $U \to 0$ and $k = 1$.

The left and the right sides of (1) will be equal if in this case

$$\delta \to -\frac{1}{2} + \Phi(U). \tag{23}$$

As it follows from (20), $\Phi(U) \to \infty$ means that $|\delta_1(U)| \to \infty$. Note that $|\delta_1| \to \infty$ does not mean that $\delta \to \infty$.

A simple expression satisfying all the above conditions a)-d) is

$$\delta(U) = \frac{2\delta_1(U)}{1 + t + \sqrt{(1-t)^2 + 16\delta_1^2(U)}} \tag{24}$$

$$t = 8\Phi(U)|\delta_1(U)|k(r) - 1 \tag{25}$$

For small $|\delta| \sim N^{-1}$ we obtain from (24)

$$\delta(U) = \delta_1(U) - 4\delta_1^3(U)[1+t] + O(t^2/N^3) \tag{26}$$

in agreement with (13). What we can say about the exactness of (1) with $\delta$ (24)? Since (24) is exact in both limiting cases $\delta_1 \to 0$ i.e. $N \to \infty$ and $\delta_1 \to \infty$ i.e.

$N \to 1$ (if $k = 1$), we evaluate an additional shift in the right side of (1) which is not taken into account, as $O(t^2)$ independently on $\beta$.

If we do not know $\delta(U)$, we can use a simpler expression similar to (24), putting $t = 0$ and expressing $\delta(U)$ through $\Phi(U)$ accordingly to (20) with an error $O(t)$.